\documentclass[prd,preprint,superscriptaddress,preprintnumbers,eqsecnum,showpacs,nofootinbib,nobibnotes,noeprint]{revtex4-1}

\usepackage{amssymb}
\usepackage{lipsum}
\usepackage{color}
\usepackage{graphicx}
\usepackage{dcolumn}
\usepackage{bm}
\usepackage[utf8]{inputenc}
\usepackage{gensymb}
\usepackage{latexsym}
\usepackage{amsmath}
\usepackage{amsfonts}
\usepackage{nicefrac}
\usepackage{float}
\usepackage{booktabs}
\usepackage{siunitx}
\usepackage{slashed}
\usepackage{hhline}
\usepackage[table]{xcolor}
\usepackage[mathscr,scaled=1.15]{urwchancal}

\usepackage{url}
\usepackage{xspace}
\usepackage{siunitx}
\usepackage{xfrac}
\usepackage{hyperref}
\usepackage[nameinlink]{cleveref}
\usepackage{appendix}

\usepackage{xifthen}
\usepackage{xcolor}
\hypersetup{colorlinks,	linkcolor={blue},	citecolor={blue},	urlcolor={blue}}

\DeclareFontFamily{OT1}{pzc}{}
\DeclareFontShape{OT1}{pzc}{m}{it}%
{<-> s * [1.15] pzcmi7t}{}
\DeclareMathAlphabet{\mathpzc}{OT1}{pzc}{m}{it}

\newcommand{\be}{\begin{equation}}
	\newcommand{\bea}{\begin{eqnarray}}
		\newcommand{\ee}{\end{equation}}
	\newcommand{\eea}{\end{eqnarray}}
\newcommand{\sla}{\slash \hspace{-0.22cm}}

\def\1eq#1{\labelcref{#1}}
\def\ie{{\it i.e.}, }
\def\eg{{\it e.g.}, }

\def\2eqs#1#2{Eqs.~(\ref{#1}) and (\ref{#2})}
\def\s#1{{\scriptscriptstyle #1}}
\def\fig#1{\Cref{#1}}
\newcommand{\Cm}[1]{{#1}_{\text{av}}}

\def\sect#1{\hyperref[#1]{\Cref{#1}}}

\graphicspath{{./figures/}{./}}

\begin{document}

\title{Electromagnetic properties  of heavy-light mesons}

\author{A.~S.~Miramontes}
\email{angel.s.miramontes@uv.es}
\affiliation{\mbox{Department of Theoretical Physics and IFIC, University of Valencia and CSIC}, E-46100, Valencia, Spain}

\author{J.~Papavassiliou}
\email{joannis.papavassiliou@uv.es}
\affiliation{\mbox{Department of Theoretical Physics and IFIC, University of Valencia and CSIC}, E-46100, Valencia, Spain}

\author{J.~M.~Pawlowski}
\email{j.pawlowski@thphys.uni-heidelberg.de}
\affiliation{\mbox{Institut f\"ur Theoretische Physik, Universit\"at Heidelberg}, Philosophenweg 16, Heidelberg, 69120, Germany}
\affiliation{\mbox{ExtreMe Matter Institute EMMI, GSI, Planckstrasse 1, Darmstadt, 64291, Germany}}

\begin{abstract}
    Within the Bethe-Salpeter framework, we present a computation of space-like electromagnetic form factors for pseudoscalar mesons, including light and heavy-light systems.  Our approach employs a flavour-dependent 
    variation of the standard Taylor effective charge, 
    which contains key contributions 
    from the quark-gluon vertices. 
    This effective interaction is a common  
    ingredient of all relevant dynamical equations, and  accommodates the crucial mass differences between the various quark flavours. 
    Particular attention is paid to the nonperturbative determination of the quark-photon vertex.
    The computed electromagnetic form factors for the pion and the kaon mesons show excellent agreement with experimental determinations. In addition, the 
    predictions for the charge radii of heavy-light systems are in overall good agreement  with lattice QCD. 
\end{abstract}

\maketitle

\section{Introduction}
\label{sec:Intro}

The electromagnetic form factors (EFF) of light mesons, in particular those of the pion and the kaon, are of key importance for the understanding of the internal structure and dynamics of hadrons in terms of their quark and gluon constituents. Over the past decades, both experimental measurements and theoretical investigations have made significant progress in exploring these form factors across various momentum-transfer regimes.

From the theoretical point of view, hadronic form factors have been extensively studied within the framework of the Bethe-Salpeter equations (BSEs), a covariant approach for describing bound states in quantum field theory \cite{Salpeter:1951sz,PhysRev.84.350,Bethe1957,Nakanishi:1969ph,Jain:1993qh,Munczek:1994zz}. This framework has been adopted in numerous studies of the EFF of light mesons~\cite{Maris:1999bh, Cloet:2008re,El-Bennich:2008dhc, Sanchis-Alepuz:2013iia, Weil:2017knt, Eichmann:2019tjk, Eichmann:2019bqf, Miramontes:2021xgn, Miramontes:2021exi, Xu:2023izo, Eichmann:2017wil, Raya:2015gva, Xu:2024fun, Miramontes:2024fgo,Xu:2025sxw,Miramontes:2025ofw} and spectroscopy~\cite{Eichmann:2016yit,Maris:1999nt, Alkofer:2002bp, Nicmorus:2008vb, Hilger:2015hka, Hilger:2014nma, Mojica:2017tvh, Serna:2017nlr,Miramontes:2022mex,Gao:2024gdj, Miramontes:2025imd}, achieving progressively higher levels of precision and sophistication. On the lattice QCD side, a plethora of calculations is available for light mesons~\cite{QCDSFUKQCD:2006gmg, Frezzotti:2008dr, Boyle:2008yd, JLQCD:2009ofg, Brandt:2013dua, Meyer:2011um, Erben:2019nmx}.

On the experimental side, the precise determination of the pion and kaon EFF remains a priority for contemporary and future hadronic physics programs. Upcoming experiments at major facilities, such as Jefferson Lab~\cite{Accardi:2023chb}, the U.S. Electron-Ion Collider (EIC)~\cite{AbdulKhalek:2021gbh}, and the Electron-Ion Collider in China (EicC)~\cite{Anderle:2021wcy}, are expected to provide accurate data on light meson form factors, particularly in the high-$Q^2$ region. 

Within the BSE formalism, EFF are calculated using the so-called ``impulse approximation", where the interaction of the photon with the quark-antiquark system is mediated via the dressed quark-photon vertex and the Bethe-Salpeter (BS) amplitude of the meson. An important ingredient of this calculation is the quark-photon vertex itself, which we compute here using its full general structure. In particular, we include both the longitudinal components satisfying the Ward-Takahashi identity (WTI) for electromagnetic current conservation, as well as the transverse components that capture dynamical effects beyond the bare vertex. This comprehensive treatment is essential for the accurate description of the EFF, particularly in heavy-light systems, where flavour asymmetries can enhance the impact of transverse contributions.

While the light sector has been thoroughly analyzed \cite{Maris:1999bh,Cloet:2008re,Eichmann:2008ae, El-Bennich:2008dhc, Sanchis-Alepuz:2013iia, Weil:2017knt, Eichmann:2019tjk, Eichmann:2019bqf, Miramontes:2021xgn, Miramontes:2021exi, Xu:2023izo, Eichmann:2017wil, Raya:2015gva, Xu:2024fun, Miramontes:2024fgo, Miramontes:2025ofw}, the inclusion of heavy-light systems, such as the $D$, $B$, and their strange counterparts ($D_s$, $B_s$), presents unique challenges, including the treatment of asymmetric quark masses and singularities in the quark propagators. 

In this work, we employ the BSE formalism to compute the EFF of both light and heavy-light pseudoscalar mesons. To account for the nontrivial flavour dynamics in these systems, we implement the flavour-dependent interaction previously developed 
in~\cite{Gao:2024gdj}. In this approach, the interaction strength acquires a natural dependence on the quark flavour via the dynamically generated quark wave function. By incorporating this flavour sensitivity directly into the BSE and gap equation, we can investigate how mass differences between the constituent quarks influence the momentum dependence of the EFF.

The article is organized as follows.
In \sect{sec:genfra} we present the general framework that will be employed throughout this work. In \sect{sec:awti} we present a 
quantitative study of the 
amount of symmetry violation 
stemming from the use of the flavour-dependent interaction in the quark gap equation. 
In \sect{sec:qpv} we carry out a  
detailed analysis of the quark-photon vertex, and present results for all of its form factors. 
\sect{sec:res} contains the 
main results of this work, namely the EFF 
and charge radii, together with an extensive comparison with experiment, lattice QCD, and a variety of approaches. 
Finally, in 
\sect{sec:conc} we discuss our conclusions.

\section{General framework}
\label{sec:genfra}

In this section we review the standard 
framework employed for the 
computation of the EFF, and discuss the key dynamical equations and their main ingredients.

\subsection{Electromagnetic form factors in the impulse approximation}
\label{sec:impulse}

Mesonic form factors are extracted from the physical amplitude that describes the interaction between a meson and an  
electromagnetic current, 
$j_{\mu}(x) = {\overline \psi}(x) \gamma_{\mu} \psi(x) $.
In the BSE framework, the current is calculated by means of the coupling of an external photon to each of the constituents of the bound state. The corresponding hadronic matrix element is given by $J_{\mu}(\Cm{p},q) = \langle\textbf{s}(p_f)|j_{\mu}|\textbf{s}(p_i)\rangle$, where $\textbf{s}(\Cm{p})$ denotes the meson under consideration. This matrix element is related to the electromagnetic form factor $F_{\textbf{s}}$  
of the meson through the expression
\begin{equation}
J^{\mu}(\Cm{p},q) = 2\Cm{p}^{\mu} \,F_{\textbf{s}} (q^2)\, \longrightarrow \, F_{\textbf{s}} (q^2) = \frac{ J\!\cdot\!\Cm{p}}{2\Cm{p}^2}
\,,
\label{current}
\end{equation}
where $\Cm{p}=(p_f+p_i)/2$ is the total momentum of the meson, $p_i$ and $p_f$ 
denote the initial and final meson momentum, respectively, 
and $q=p_f-p_i$ is the photon momentum.

\begin{figure*}[t]
\centerline{%
\includegraphics[width=1.0\textwidth]{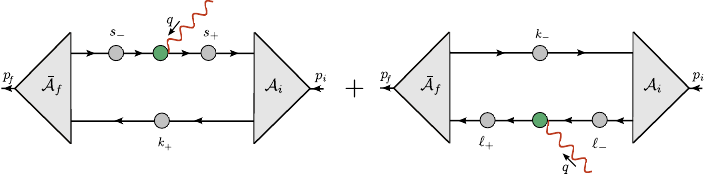}}
\caption{Diagrammatic representation of electromagnetic current in the impulse approximation, employed for the calculation of electromagnetic form factor. The green and gray circles denote the fully-dressed quark-photon vertex and quark propagator,
respectively, and the 
various kinematic variables are defined in \1eq{eq:kinematics}.}
\label{fig:currentJ}     
\end{figure*}

In the impulse approximation \cite{Eichmann:2011aa, Eichmann:2016yit},  
the conserved current $J^{\mu}(\Cm{p},q)$ describes the coupling of a single photon to a quark-antiquark system, see  \Cref{fig:currentJ}. 
The $J^{\mu}(\Cm{p},q)$
is given by the following integral expression 
\begin{equation}
J^{\mu}(\Cm{p},q) = \int_k \bar{\mathcal{A}}(k_f, p_f) S (\ell_{+}) \Gamma^{\mu}(k_+,q) S (\ell_{-}) \mathcal{A}(k_i, p_i) S (k_{-}),
\label{eq:current}
\end{equation}
where $\mathcal{A}$ is the BS amplitude and $\bar{\mathcal{A}}$ its  charge conjugated counterpart,  $\Gamma^{\mu}$ represents the quark-photon vertex, and $S$ stands for the quark propagator. 
The integral measure is denoted by %
\begin{align}
\int_{k} := (2\pi)^{-4} \int_{-\infty}^{+\infty} \!\!{\rm d}^4 k \,,
\end{align}
where the use of a symmetry-preserving regularization scheme is implicitly assumed. In addition, the relevant kinematic variables are defined as 
\begin{align}
&k_+ = k + \eta \Cm{p}\,, \qquad k_- = k - (1-\eta)\Cm{p},\qquad 
\ell_+ = k_+ + \frac{q}{2},  \nonumber\\
& \ell_- = k_+ - \frac{q}{2} \,, \qquad \qquad s_+ = k_- + \frac{q}{2}, \qquad \qquad s_- = k_- - \frac{q}{2} \,.
\label{eq:kinematics}
\end{align}
Here, the parameter $\eta$ determines the distribution of the total momentum $\Cm{p}$ between the quark and antiquark. 
Depending on the quark mass composition, a suitable choice  
of the value of $\eta$ 
leads to a tighter control  of the  singularities in the quark propagator.
This, in turn, 
enhances the numerical stability of the calculation, especially  for systems with significant mass disparity, such as heavy-light mesons.

The processes under consideration 
are elastic, 
having the same initial and final states, whose 
common bound state mass is denoted by $M$. Thus,
the on-shell momenta satisfy 
\mbox{$p_i^2 = p_f^2=-M^2$}, and  $\Cm{p}$ 
may be parametrized as 
$\Cm{p} = ( 0 , 0 , 0 , i\sqrt{ M^2 + q^2/4})$.

It is important to note that \1eq{eq:current}
accounts for the case where the photon interacts with only one of the valence quarks of the meson; evidently, 
the contribution of the interaction of the photon with the other quark 
must be duly added, as shown 
in \fig{fig:currentJ}. Thus, the complete electromagnetic form factor for a pseudoscalar particle reads,
\begin{equation}
    F_{\textbf{s}}(q^2)= e_q F_{q}(q^2)+ e_{\overline{q}} F_{\overline{q}}(q^2)
    \,,
    \label{eq:FF_full}
\end{equation}
where $e_q$ and $e_{\overline{q}}$ are the corresponding electric charges for the quark and antiquark.

\subsection{Dynamical equations}
\label{sec:dyneq}

\begin{figure*}[t]
\centerline{%
\includegraphics[width=0.9\textwidth]{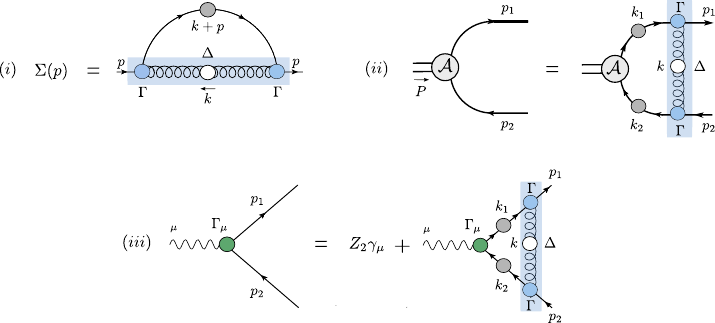}}
\caption{Diagrammatic representation of the main dynamical components: 
(${\it i}$) is the quark self-energy 
entering in the gap equation; 
(${\it ii}$) the BSE for the mesonic wave function; and 
(${\it iii}$) the SDE of the quark-photon vertex. 
Appropriate contributions from the quark-gluon vertices (blue circles) are combined with the 
scalar factor, $\Delta(k)$, of the Landau-gauge gluon propagator to form 
the interaction kernel 
of \1eq{eq:Cfun}, depicted here by the blue band.}
\label{fig:vertices}     
\end{figure*}

The elements comprising 
\1eq{eq:current},
namely $S$, ${\cal A}$, 
and $\Gamma^{\mu}$
are determined 
from their own dynamical equations, 
evaluated within appropriately constructed truncation schemes. 
Recently, a new framework 
for the study of heavy-light mesons  
has been put forth~\cite{Gao:2024gdj}, where the standard 
one-gluon exchange interaction is complemented by 
flavour-dependent contributions 
stemming from 
the adjacent quark-gluon vertices
(blue circles in \fig{fig:vertices}).
In this treatment, 
only the classical form factor of each quark-gluon vertex is retained, and is evaluated
in the so-called “symmetric” configuration. The Slavnov-Taylor identity satisfied by the quark-gluon vertex links this form factor to the quark wave-function, $A_f (q^2)$, see \1eq{eq:qprop}, 
which encodes the flavour-dependence. 
Specifically, the resulting 
effective interaction, 
${\mathcal I}_{ff'}(q^2)$, is given by 
\begin{equation}
{\mathcal I}_{ff'}(q^2)
 = 
{\tilde\alpha}_{T}(q^2) A_{f}(q^2) 
A_{f'}(q^2) \,,
\label{eq:Cfun}
\end{equation}
where the effective charge 
${\tilde\alpha}_{T}(q^2)$ is the so-called ``modified Taylor coupling", introduced 
in~\cite{Gao:2024gdj}. 
The difference 
between ${\tilde\alpha}_{T}(q^2)$ and the standard 
Taylor coupling, ${\alpha}_{T}(q^2)$, is the 
inclusion of certain 
process-independent contributions extracted from the quark-gluon vertex. 
A useful fitting function describing 
${\tilde\alpha}_{T}(q^2)$ is given by 
Eq.~(20) of~\cite{Gao:2024gdj}. 

The effective interaction 
of \1eq{eq:Cfun} is a common 
ingredient of the three main dynamical equations, namely
(\ref{eq:gapI}), (\ref{eq:BSEI}) and (\ref{eq:vecvert});  
in fact, in all of them, it appears in the unique combination
\begin{align}
D_{ff'}^{\mu\nu}(k):= 4 \pi\Delta_0^{\mu\nu}(k) 
\,
{\mathcal I}_{\!ff'}(k^2) \,,
\label{eq:theD}
\end{align}
with
\begin{align}
\Delta_0^{\mu\nu}(k) = 
\left(\delta^{\mu\nu} - 
\frac{k^{\mu}k^{\nu}}{k^2}\right)\frac{1}{k^2}\,.
\label{eq:gluonprop}
\end{align}
the tree-level gluon propagator in the Landau gauge. 

In particular, we have : 

(${\it i}$)
The 
quark propagator, \mbox{$S_{\!f}^{ab}(p)=i\delta^{ab}S_{\!  f}(p)$}, where the index $f$ stands for the quark flavour, 
taking values 
\mbox{$f=u,d,s,c,b$}.
The standard decomposition 
of $S_{\!f}^{-1}(p)$
is 
\begin{align} 
\label{eq:qprop}
    S_{\!f}^{-1}(p)=i \slashed{p} \, A_{f}(p^2)+B_{f}(p^2)\,,
\end{align}
where $A_{f}(p^2)$ and $B_{f}(p^2)$ are the dressings of the Dirac vector and scalar tensor structures,  respectively. The renormalization-group invariant
quark mass function, 
${\mathcal M}_{f}(p^2)$,  
is given by \mbox{${\mathcal M}_{f}(p^2) = B_{f}(p^2)/ A_{f}(p^2)$}.
At tree-level, 
$S_{0,f}^{-1}(p)= i\slashed{p} + m_{f}$\,, 
where $m_{f}$ denotes the current quark mass of the flavour $f$. In addition, the  
 {\it self-energy}, 
 $\Sigma_f(p^2)$, 
is 
 defined as \mbox{$\Sigma_{f}(p^2) = S_{\!f}^{-1}(p) - S_{0,f}^{-1}(p)$}.

The evolution of the 
components $A_{f}(p^2)$ and $B_{f}(p^2)$ is determined from the quark gap equation. In terms of the kernel introduced in~\cite{Gao:2024gdj}, the renormalized gap equation reads 
\begin{align}
S_{\!f}^{-1}(p) = 
Z_{2}(i\sla{p}+m_{\!\s R}) + 
\underbrace{C_{\!\s F} \!\!\int_k\,
\gamma_{\alpha} S_{\!f}(k+p)\gamma_{\beta} 
\,D_{ff}^{\alpha\beta}(k)}_{\Sigma_f(p^2)} \,,
\label{eq:gapI}
\end{align}
where
\mbox{$C_{\!\s F}=4/3$} is the Casimir eigenvalue of the fundamental representation, and $m_{\!\s R}$ is the renormalized current quark mass. 
The diagrammatic 
representation of the quark self-energy 
$\Sigma_f(p^2)$ is shown in 
\fig{fig:vertices}, panel (${\it i}$).
Finally, 
$Z_2$ is the wave-function renormalization of the quark field, determined within the 
momentum subtraction (MOM) renormalization scheme \cite{Celmaster:1979km, Hasenfratz:1980kn, Braaten:1981dv, Athenodorou:2016oyh}.

(${\it ii}$)
The BS amplitude $\mathcal{A}$ in \1eq{eq:current} is given by the homogeneous BSE
[see 
panel (${\it ii}$)] of 
\fig{fig:vertices})
\begin{align}
{\cal A}_{ff'}(p_1,p_2) = 
-\int_{k} \gamma_{\mu}
S(k_1) {\cal A}_{ff'}(k_1,k_2) S(k_2)
\gamma_{\nu} D_{ff'}^{\mu\nu}(k)
 \,,
\label{eq:BSEI}
\end{align}
with $k_i = k + p_i, \,\,i =1,2$.

An alternative parametrization 
of the momenta, used extensively in the related literature, is given by 
$P = p_1-p_2$ and 
$p = (p_1+p_2)/2$. 
In terms of these variables,
the tensorial decomposition of the BS amplitude of a pseudoscalar meson is given by 
\begin{align}
     \mathcal{A}_{ff'}(p,P) = 
     \left( \chi_{\s 1}^{\s f \s f'} +i \chi_{\s 2}^{\s f \s f'} \!\slashed{P} + i \chi_{\s 3}^{\s f \s f'} \!\slashed{p}(p \cdot P) + \chi_{\s 4}^{\s f \s f'} [\slashed{p}, \slashed{P}] \right) \gamma_5
     \,,
\label{eq:thechis}     
 \end{align} 
where the subamplitudes 
$\chi_i^{\s f \s f'} := \chi^{\s f \s f'}_i(p,P)$
are functions of the 
Lorentz scalars $P^2$, $p^2,p\cdot P$.

(${\it iii}$)
The SDE of the quark-photon vertex,
depicted diagrammatically in
panel (${\it iii}$) of 
\fig{fig:vertices}, 
is given by 
\begin{align}
\Gamma^{f}_{\!\mu}(p_1, p_2) = Z_2\gamma_{\mu} - 
C_{\!\s F} \!\!\int_k 
\gamma_{\alpha}
S_{\!f}(k_1) \Gamma_{\!\!\mu}^{f}(k_1,k_2)S_{\!f}(k_2)
 \, 
\gamma_{\beta} 
D_{ff}^{\alpha\beta}(k)\,. 
\label{eq:vecvert}
\end{align}
The self-consistent treatment of \1eq{eq:vecvert} is 
presented in the next section,
where the momentum-dependence  of the 
twelve form-factors comprising $\Gamma^{f}_{\!\mu}(p_1, p_2)$ will be determined.

\section{Axial Ward-Takahashi identity and the effective kernel}
\label{sec:awti}

Of central importance for  
the dynamical chiral symmetry breaking are the 
axial-vector current, 
$j^{a}_{5\mu}(x)$,
and the axial current,
$j^{a}_{5}(x)$, see,\eg\!\cite{Itzykson:1980rh} (Ch.11), and~\cite{doi:10.1142/2170}.
In terms of the fundamental quark fields, $q(x)$ and $\bar{q}(x)$, they are 
defined as  
$j^{a}_{5\mu}(x) = \bar{q}(x)\gamma_5 \frac{\lambda^{a}}{2}\gamma_{\mu}q(x)$
and 
$j^{a}_{5}(x) = \bar{q}(x)\gamma_5 \frac{\lambda^{a}}{2} q(x)$, where 
$\lambda^{a}$ are the Gell-Mann matrices. 
Then, the axial-vector vertex, 
$\Gamma^{a}_{5\mu}(p_1, p_2)$,
and the axial vertex, 
$\Gamma^{a}_{5}(p_1, p_2)$,
are defined as the momentum-space 
transforms of 
$\langle 0\vert T [j^{a}_{5\mu}(0)q(x)\bar{q}(y)]\vert 0\rangle$
and $\langle 0\vert T [j^{a}_{5}(0)q(x)\bar{q}(y)]\vert 0\rangle$,
respectively, where $T$ denotes the standard time-ordering operator. 

\begin{figure*}[t]
\centerline{%
\includegraphics[width=1.0\textwidth]{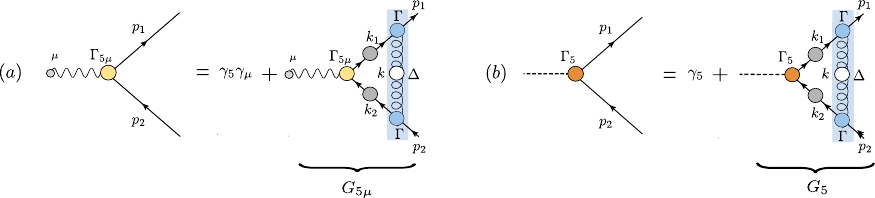}}
\caption{Diagrammatic representation of the 
SDE governing the vertex $\Gamma_{5\mu}^{ff'}(p_1, p_2)$ [panel $(a)$], and of the 
SDE governing the vertex $\Gamma_{5}^{ff'}(p_1, p_2)$ [panel $(b)$]. 
The wavy line with a small gray circle at its end denotes 
the axial-vector current $j^{a}_{5\mu}(x)$, while the 
dotted line indicates the 
axial current $j^{a}_{5}(x)$. 
As in the case of \fig{fig:vertices}, 
the interaction kernel (blue band) is given by \1eq{eq:Cfun}.}
\label{fig:axial_vertices}     
\end{figure*}

We next suppress the “isospin" 
indices, and 
introduce quark flavour indices, 
$f$ and $f'$. Then, it is 
well-known that 
the axial-vector vertex
$\Gamma_{5\mu}^{ff'}(p_1, p_2)$ 
satisfies the WTI \cite{Itzykson:1980rh, doi:10.1142/2170}
\begin{align}
P^{\mu} \Gamma_{5\mu}^{ff'}(p_1, p_2) =  S^{-1}_{f}(p_1)  i\gamma_5 +  i\gamma_5 S^{-1}_{f'}(p_2)
- i(m_f + m_{f'})\Gamma_{5}^{ff'}(p_1, p_2)  \,,
\label{eq:AWTI} 
\end{align}
where $P=p_1-p_2$. 

Within the framework of~\cite{Gao:2024gdj}, 
the vertices 
$\Gamma_{5\mu}^{ff'}(p_1, p_2)$ and 
$\Gamma_{5}^{ff'}(p_1, p_2)$
satisfy the SDEs shown in 
\fig{fig:axial_vertices}, namely 
\begin{align}
\Gamma^{ff'}_{\!5 \mu}(p_1, p_2) = \gamma_5\gamma_{\mu} 
\,\underbrace{- 
\int_k 
\gamma_{\alpha}
S_{\!f'}(k_1) \Gamma_{5\mu}^{ff'}(k_1,k_2)S_{\!f}(k_2)
 \, 
\gamma_{\beta} 
D_{ff'}^{\alpha\beta}(k)}_{G^{ff'}_{\!5 \mu}(p_1, p_2) }\,, 
\label{eq:axial}
\end{align}
and 
\begin{align}
\Gamma_{5}^{ff'}(p_1, p_2)
= \gamma_5 \,
\underbrace{- 
\int_k 
\gamma_{\alpha}
S_{\!f'}(k_1) \Gamma_{5}^{ff'}(k_1,k_2)S_{\!f}(k_2)
 \, 
\gamma_{\beta} 
D_{ff'}^{\alpha\beta}(k)
}_{G^{ff'}_{5}(p_1, p_2) }\,,
\label{eq:gamma5}
\end{align}
where, for latter convenience, 
we use the short-hand notation 
$G^{ff'}_{\!5 \mu}(p_1, p_2)$ 
and $G^{ff'}_{\!5}(p_1, p_2)$ 
to denote the non-trivial parts of the corresponding SDEs. 

Evidently, upon contraction 
with $P^{\mu}$, 
the tree-level term  $\gamma_5 \gamma_{\mu}$
yields 
\begin{align}
P^{\mu} \gamma_5 \gamma_{\mu}
= S_{0,f}^{-1}(p_1) i\gamma_5 + i \gamma_5 S_{0,f'}^{-1}(p) -2 i (m_f+m_{f'})
\Gamma_{0,5}^{ff'}(p_1, p_2)\,,
\label{eq:WI0}
\end{align}
where we have used that, at tree-level, $\Gamma_{0,5}^{ff'}(p_1, p_2) =
\gamma_5$. 

Turning to the quantum part of the calculation,
one may directly confirm that, 
in the case of equal flavours,  
$f=f'$, the use of the kernel 
$I_{ff}(q^2)$ in \1eq{eq:axial}
and \1eq{eq:gamma5} preserves the 
validity of \1eq{eq:AWTI}. 
In particular, 
the contraction of  
$\Gamma_{5\mu}^{ff'}(k_1, k_2)$
by $P^{\mu}$ triggers 
the WTI of \1eq{eq:AWTI}
under the integral sign of  
$G^{ff}_{\!5 \mu}(p_1, p_2)$,
and one gets 
\begin{align}
P^{\mu}
G^{ff}_{\!5 \mu}(p_1, p_2)
=&\, \!\!- i 
\!\int_k 
\gamma_{\alpha}
S_{\!f}(k_1) 
[S^{-1}_{f}(k_1)  \gamma_5 +  \gamma_5 S^{-1}_{f}(k_2)
-2 m_f \Gamma_{5}^{ff}(k_1, k_2)]
S_{\!f}(k_2)
 \, 
\gamma_{\beta} 
D_{ff}^{\alpha\beta}(k)
\nonumber\\[1ex]
=&\, 
i \gamma_5  \!\int_k 
\gamma_{\alpha}
S_{\!f}(k_2)
 \, 
\gamma_{\beta} 
D_{ff}^{\alpha\beta}(k)
+
i \!\int_k 
\gamma_{\alpha}
S_{\!f}(k_1) 
\gamma_{\beta} 
D_{ff}^{\alpha\beta}(k)
\gamma_5
\nonumber\\[1ex]
& \,+ 
2i m_f 
\int_k 
\gamma_{\alpha}
S_{\!f'}(k_1) \Gamma_{5}^{ff'}(k_1,k_2)S_{\!f}(k_2)
 \, 
\gamma_{\beta} 
D_{ff}^{\alpha\beta}(k)
\,,
\nonumber\\[1ex]
=&\, 
\Sigma_f (p_1) i\gamma_5 + 
i\gamma_5  \Sigma_f (p_2)  
- 2 i m_f G_{5}^{ff}(p_1, p_2) \,.
\label{eq:WI1} 
\end{align}
Then, the sum of 
\1eq{eq:WI0} and \1eq{eq:WI1} 
yields precisely 
the two sides of 
\1eq{eq:AWTI}, with 
$f=f'$.

Instead, 
when $f \neq f'$,
the use of the kernel 
$I_{ff'}(q^2)$ leads to a 
violation of \1eq{eq:AWTI}.
In particular, the repetition of the steps 
leading to \1eq{eq:WI1} 
now yields 
\begin{align}
P^{\mu}
G^{ff'}_{\!5 \mu}(p_1, p_2) = 
\widetilde\Sigma_f (p_1) i\gamma_5 + 
i\gamma_5  \widetilde\Sigma_{f'} (p_2)  
- i (m_f+m_{f'}) G_{5}^{ff'}(p_1, p_2) \,,
\label{eq:WI2} 
\end{align}
where
\begin{align}
\widetilde\Sigma_f (p_1) =  C_{\!\s F} \!\!\int_k\,
\gamma_{\mu} S_{\!f}(k_1)\gamma_{\nu} 
D_{ff'}^{\mu\nu}(k)
 \,,
\qquad
\widetilde\Sigma_f (p_2) =  C_{\!\s F} \!\!\int_k\,
\gamma_{\mu} S_{\!f}(k_2)\gamma_{\nu} 
D_{ff'}^{\mu\nu}(k) \,.
\label{eq:mixprop}
\end{align}
We note that, in contradistinction to 
$\Sigma_f (p_1)$ and $\Sigma_f (p_2)$,
the 
$\widetilde\Sigma_f (p_1)$
and $\widetilde\Sigma_f (p_2)$ in \1eq{eq:mixprop}
contain a non-diagonal
$D_{ff'}^{\mu\nu}(k)$, which thwarts 
the interpretation of these 
quantities as genuine self-energies. 
Evidently, this mismatch  
is the manifestation of the  
WTI violation associated with 
the use of the 
kernel $I_{ff'}(q^2)$
when  
$f\neq f'$.

\begin{figure*}[t]
\centerline{%
\includegraphics[width=0.9\textwidth]{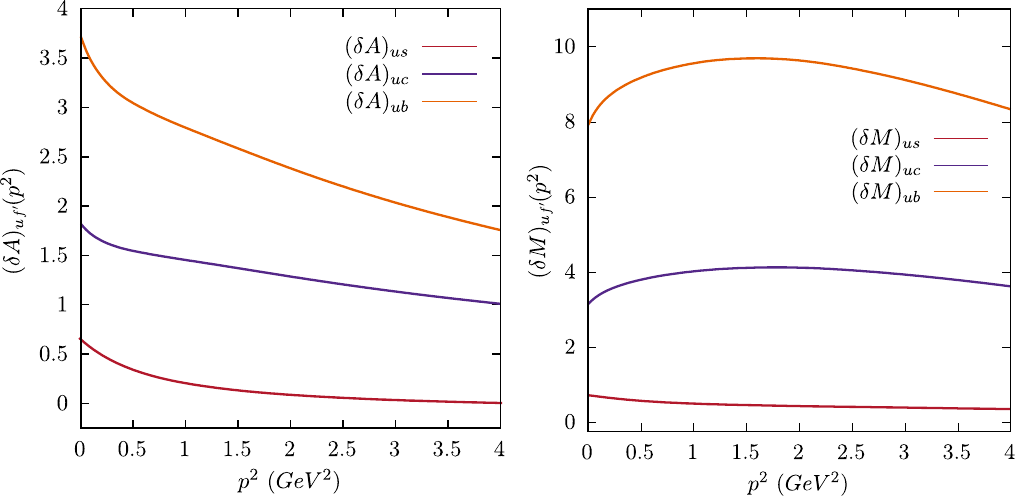}}
\caption{The relative errors.}
\label{fig:delta_mass}     
\end{figure*}

A way to quantify this violation is to 
study the quantitative difference between 
the components of the $S_f(p)$ computed using 
$\Sigma_f (p)$,  and the same $S_f(p)$ computed using 
$\widetilde\Sigma_f (p)$ for all possible $I_{ff'}(q^2)$.
In particular, 
if we fix $f=u$, we may first compute $\Sigma_u (p)$
using  the correct kernel $I_{uu}$; 
we will denote the resulting quark propagator by 
$\Sigma_{uu} (p)$, 
where the additional subscript ``$u$" indicates 
precisely that $I_{uu}$ was used in the computation. 
Then, we compute $\Sigma_u (p)$ using instead 
$\widetilde\Sigma_u (p)$, which, depending on the value of $f'$, 
may contain $I_{us}$, $I_{uc}$, or $I_{ub}$; this gives rise to 
three versions of $\Sigma_u (p)$, which we denote by 
$\Sigma_{us} (p)$, $\Sigma_{uc}(p)$, and 
$\Sigma_{ub}(p)$, respectively. 
Employing the standard decomposition given in \1eq{eq:qprop}, 
we will denote the 
corresponding Dirac components of the above  
quark propagators by $A_{uu}(p^2)$, $A_{uf'}(p^2)$, and 
$B_{uu}(p^2)$, $B_{uf'}(p^2)$, and the resulting 
versions of the $u$ quark mass by 
${\mathcal M}_{uu}(p^2)$ and 
${\mathcal M}_{uf'}(p^2)$, 
with $f'=s,c,b$. 

With the above ingredients in hand, we 
next introduce the 
point-wise percentage errors 
$(\delta A)_{uf'}(p^2)$
and 
$(\delta \mathcal M)_{uf'}(p^2)$, defined as 
\begin{align}
(\delta A)_{uf'}(p^2) =&\, 
\frac{2 \,\lvert A_{uu} (p^2) - A_{uf'}(p^2)\rvert}{A_{uu} (p^2) + A_{uf'}(p^2)} 
\times 100 \% \,,
\nonumber\\
(\delta \mathcal M)_{uf'}(p^2) =&\, 
\frac{2 \,\lvert \mathcal M_{uu} (p^2) - \mathcal M_{uf'}(p^2)\rvert}{\mathcal M_{uu} (p^2) + \mathcal M_{uf'}(p^2)} 
\times 100 \% \,,
\label{eq:errors}
\end{align}
for $f'=s,c,b$.
The curves for 
$(\delta A)_{uf'}(p^2)$ and 
$(\delta \mathcal M)_{uf'}(p^2)$, 
obtained from the appropriate treatment of the 
gap equation,  
are shown in \fig{fig:delta_mass}.
We see that in the worst case, involving the quarks with the largest mass disparity, 
$f=u$ and $f'=b$, 
the maximum 
error in the computation of the quark wave-function is 
slightly over $3\%$ (left panel).   
The maximum discrepancy 
in the computation of the 
constituent quark masses occurs again for the case $f=u$ and $f'=b$,
and is about $9\%$ (right panel).

As an illustration, we computed the mass of the heavy light meson $D$ using propagators for both the $u$ and $c$ quarks constructed with $\Sigma_{uc}(p)$, \ie both dressed using the off-diagonal kernel $I_{uc}$. This yields a mass of $1.99~\mathrm{GeV}$, compared to the $1.93~\mathrm{GeV}$ obtained in our previous study using diagonal kernels for the propagators \cite{Gao:2024gdj}.

\section{Quark-Photon vertex}
\label{sec:qpv}

The quark-photon vertex is a crucial ingredient in the study of the electromagnetic interaction of hadrons \cite{Sanchis-Alepuz:2013iia, Eichmann:2016yit,Xu:2024fun, Miramontes:2025ofw}, and has been studied in isolation in a series of works \cite{Frank:1994mf, Maris:1999bh,  Miramontes:2019mco, Leutnant:2018dry, Tang:2019zbk}.

Just as the 
electron-photon 
vertex known from QED, 
the quark-photon vertex $\Gamma^{f}_{\!\mu}(p_1, p_2)$ satisfies 
the fundamental WTI 
\begin{align}
    q^{\mu} \Gamma^{f}_{\!\mu}(p_1, p_2)  = iS_{\!f}^{-1}(p_2) - iS_{f}^{-1}(p_1) \,.
    \label{eq:VWTI}
\end{align}
It is now straightforward to establish the compatibility of the
kernel $I_{ff}(q^2)$  
with the local $U(1)$ symmetry, 
encoded in  
the WTI of \1eq{eq:VWTI}. 
At the level of the bare vertex SDE
($Z_2=1$), one contracts by $q^{\mu}$
both sides of 
\1eq{eq:vecvert}, 
and uses 
the r.h.s. of 
\1eq{eq:VWTI} under the integral sign of \1eq{eq:vecvert}, namely 
\begin{align}
q^{\mu} \Gamma^{f}_{\!\mu}(p_1, p_2) =&\, 
\sla{p}_1- \sla{p}_2
- 
i C_{\!\s F} \!\!\int_k 
\gamma_{\alpha}
S_{\!f}(k_1) [S_{f}^{-1}(k_2) - S_{f}^{-1}(k_1)]S_{\!f}(k_2)
 \, 
\gamma_{\beta} 
D_{ff}^{\alpha\beta}(k)
\nonumber\\[1ex]
=&\,
i \left[i\sla{p}_2 + m_f + C_{\!\s F} \!\!\int_k 
\gamma_{\alpha} 
S_{\!f}(k_2) \gamma_{\beta} 
D_{ff}^{\alpha\beta}(k)
\right] 
\nonumber\\[1ex]
&\,
i \left[i\sla{p}_1 + m_f + C_{\!\s F} \!\!\int_k 
\gamma_{\alpha} 
S_{\!f}(k_1) \gamma_{\beta} 
D_{ff}^{\alpha\beta}(k) \right] \,,
\label{eq:VWTI2}
\end{align}
which, by virtue of \1eq{eq:gapI}, is precisely the WTI of 
\1eq{eq:VWTI}. 
The preservation of the $U(1)$ gauge symmetry ensures charge conservation, which is crucial for the veracious calculation of hadronic form factors.
Note, in addition, that the requirement that the WTI of 
\1eq{eq:VWTI} be valid {\it after} renormalization imposes the well-known relation $Z_1=Z_2$, where 
$Z_1$ denotes the renormalization constant of the quark-photon vertex;
this equality has already been enforced at the level of the vertex SDE in \1eq{eq:vecvert}, 
where $Z_1$ has been replaced 
by $Z_2$. 

The vertex $\Gamma_{\!\mu}(p_1, p_2)$ may be naturally decomposed into two parts, 
\begin{align}
\Gamma^{\mu}(p_1, p_2) = 
\Gamma^\mu_{\!\rm {\s B\s C}}(p_1,p_2)
+\Gamma^\mu_{\!\rm \s T}(p_1,p_2) \,,
\label{eq:verdec}
\end{align}
where the so-called ``Ball-Chiu"
part, 
$\Gamma^\mu_{\!\rm {\s B\s C}}(p_1,p_2)$, 
saturates the WTI of \1eq{eq:VWTI}, while the 
transverse part, 
$\Gamma^\mu_{\!\rm \s T}(p_1,p_2)$,
satisfies 
$q_{\mu}\Gamma^\mu_{\!\rm \s T}(p_1,p_2)=0$.
According to the 
construction of~\cite{Ball:1980ax},  
the form factors comprising 
$\Gamma^\mu_{\!\rm {\s B\s C}}(p_1,p_2)$
are expressed entirely in terms of the quark functions 
$A(p^2)$ and $B(p^2)$ appearing in \1eq{eq:qprop}. In particular, 
introducing for convenience the 
kinematic variables 
 $u=(p_1+p_2)/2$ and $q=p_1-p_2$, 
we have 
\begin{align}
\Gamma^\mu_{\!\rm {\s B\s C}}(q,u) = \lambda_1\gamma^\mu + 2 \lambda_2 \, u^\mu \slashed{u} + 2 i\lambda_3 \, u^\mu + i\lambda_4 [\gamma^\mu,\slashed{u}]  \,,  
\label{eq:vertex_expansion1}
\end{align}
with the form factors given by 
\begin{align}
\lambda_1 = \frac{A(p_1^2)+A(p^2_2)}{2} \,,\quad
\lambda_2 = \frac{A(p^2_1)-A(p^2_2)}{p^2_1-p^2_2} \,, \quad
\lambda_3 = \frac{B(p^2_1)-B(p^2_2)}{p^2_{1}-p^2_{2}} \,, \quad
\lambda_4 = 0 \,.
\label{eq:Ball_Chiu_FF}
\end{align}
The transverse component 
$\Gamma^\mu_{\!\rm \s T}(p_1,p_2)$
may be expanded as  
\begin{align}
\Gamma^\mu_{\!\rm \s T}(q,u) = \sum_{i=1}^{8} h_i(q,u) \,\tau_i^\mu, 
\label{exp}
\end{align}
where the $\tau_i^\mu$ are 
the elements of an eight-dimensional basis, and the  
$h_i(q,u)$ denote the 
associated form factors. 
Specifically, we employ the basis of~\cite{Eichmann:2014qva}, 
whose elements are given by 
\begin{figure*}[t]
\centerline{%
\includegraphics[width=1.2\textwidth]{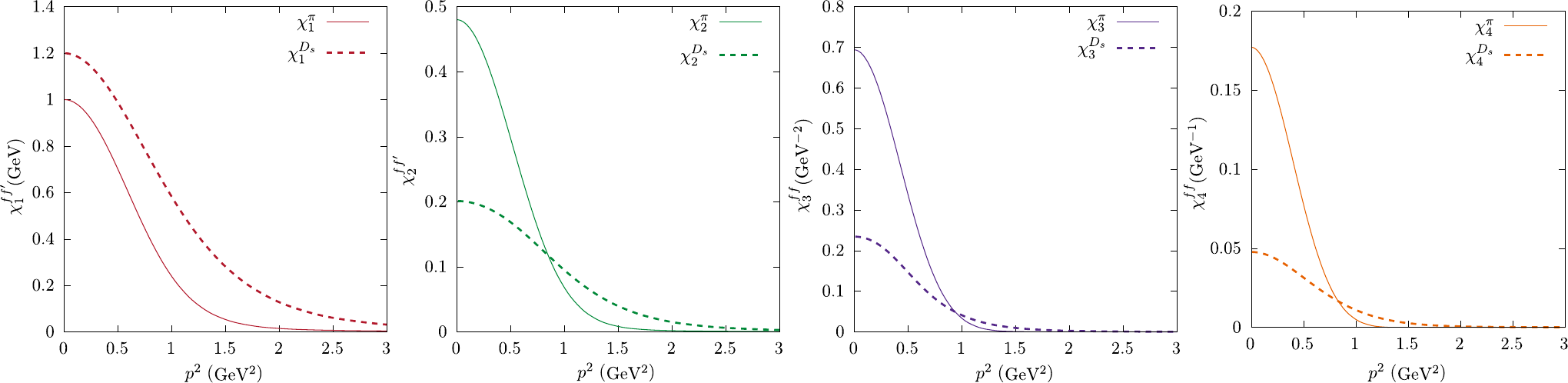}}
\caption{The BS amplitude for the pion (solid lines) and for the $D_s$ (dashed lines). We have set  $\chi_i^{\pi} : = \chi_i^{u \bar{d}}$ and $\chi_i^{D_s} : = \chi_i^{c \bar{s}}$,
where the general 
amplitude $\chi_i^{ff'}$, 
indicating the specific quark content, has been 
introduced in \1eq{eq:thechis}. For easier comparison, all BS amplitudes are rescaled such that $\chi_1^{\pi}(0) = 1.$}
\label{fig:BSA_amps}     
\end{figure*}
\begin{eqnarray}
\tau_1^\mu &=& t^{\mu\nu}_{qq} \gamma_\nu\;, \hspace{2.5cm}  
\tau_2^\mu = t^{\mu\nu}_{qq} \left(u\cdot q\right) \frac{i}{2} [\gamma_\nu,\slashed{u}]\;, \nonumber\\[1ex]
\tau_3^\mu &=& \frac{i}{2} [\gamma^\mu,\slashed{q}]\;, \hspace{1.95cm}  
\tau_4^\mu = \frac{1}{6} [\gamma^\mu,\slashed{u},\slashed{q}]\;, \nonumber\\[1ex]
\tau_5^\mu &=& i t^{\mu\nu}_{qq} u_\nu\;, \hspace{2.4cm}  
\tau_6^\mu = t^{\mu\nu}_{qq} u_\nu\slashed{u}\;, \nonumber\\[1ex]
\tau_7^\mu &=& t^{\mu\nu}_{qu} \left(u\cdot q\right) \gamma_\nu\;, \hspace{1.25cm}  
\tau_8^\mu = t^{\mu\nu}_{qu} \frac{i}{2} \left[\gamma_\nu,\slashed{u}\right]\,,
\label{eq:vertex_expansion2}
\end{eqnarray}
where we used the projector  
$t^{\mu\nu}_{ab} =\left(a\cdot b\right) \delta^{\mu\nu}-b^\mu a^\nu$, and the triple commutator 
is defined as \mbox{$[A,B,C]=[A,B]C + [B,C]A + [C,A]B$}, with 
$[A,B,A] =0$. 
It is understood that all form factors 
introduced above carry a flavour index “$f$”, which has been suppressed for simplicity.

Note that a key advantage of the decomposition 
in \1eq{eq:verdec}
is that it enables a separation of the transverse components with respect to the total momentum $q$. This is especially relevant for a time-like photon, where the quark-photon vertex develops vector meson bound-state poles (e.g., at $q^2=-M_{\rho}^2$). These poles reside exclusively in the transverse components, while the time-like domain may also feature non-resonant singularities originating from the analytic structure of the quark propagator~\cite{Williams:2018adr,Miramontes:2019mco}.

\begin{figure*}[t]
\centerline{%
\includegraphics[width=0.7\textwidth]{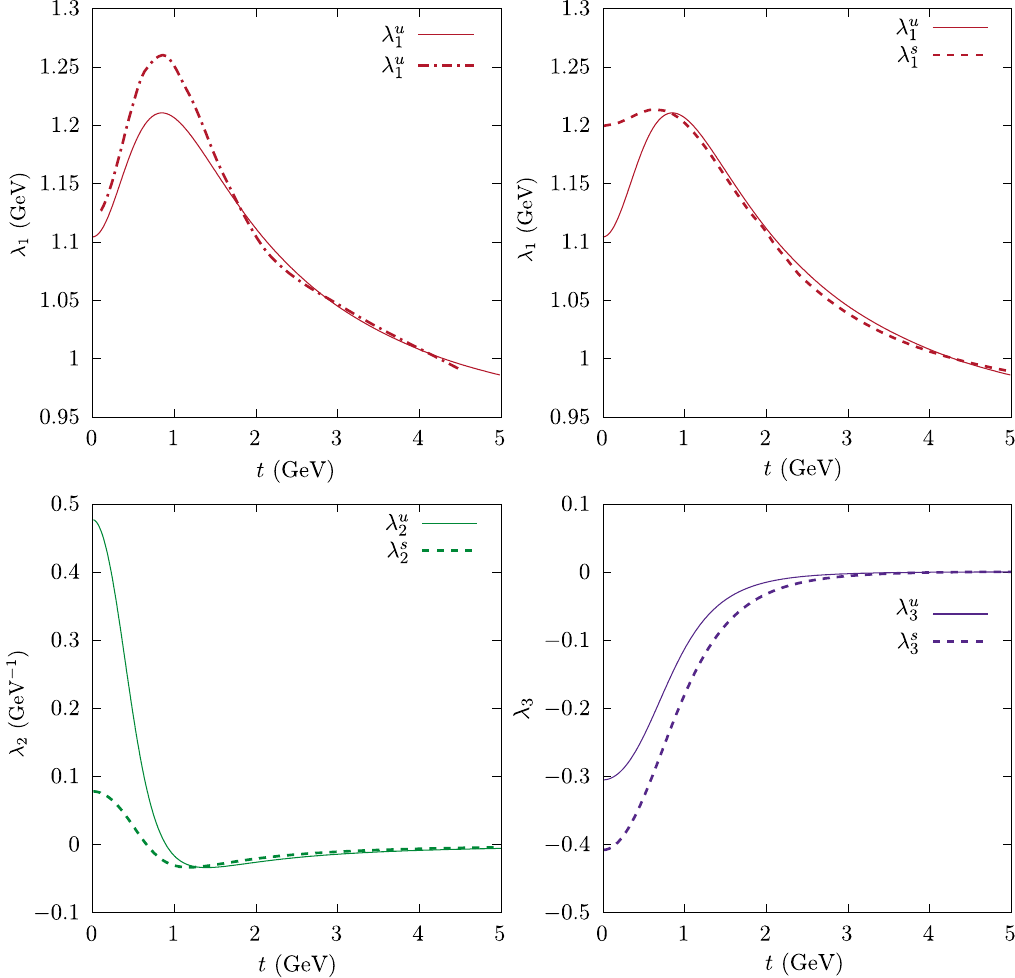}}
\caption{Longitudinal form factors of the quark-photon vertex. Upper-left: the up-quark form factors obtained using the interaction of \1eq{eq:Cfun} (solid line)  compared to the RL result of~\cite{Miramontes:2019mco}~(dash-dotted lines). Upper right: up-quark (solid line) and strange quark (dashed line) form factor $\lambda_1.$ Bottom panel: up-quark (solid line) and strange quark (dashed line) form factor $\lambda_2.$ and $\lambda_3$.}
\label{fig:qphv_dress_long}     
\end{figure*}
%

\section{Results}
\label{sec:res}

In this section we present the numerical results for the BS amplitude and quark-photon vertex. Once these elements have been obtained, the current $J_{\mu}$ can be calculated from \1eq{eq:current}, and the EFF of  pseudoscalar particles may be subsequently extracted. 

(${\it i}$) \textbf{Numerical inputs.} The current quark masses have been fixed to the following values:
\begin{align}
m_{u/d} = 0.005\,\text{GeV}, \quad m_s = 0.094\,\text{GeV}, \quad m_c = 1.1\,\text{GeV}, \quad m_b = 3.5\,\text{GeV}, 
\end{align}
at the renormalization point $\mu = 4.3$ GeV. Furthermore, the 
parameter $\eta$ that controls the routing of the total momentum $\Cm{p}$ in \1eq{eq:kinematics} has been fixed to the following values, which depend on the quark content of the meson:
\begin{align}
\eta_{us} = 0.47, \quad \eta_{cb} = 0.41, \quad \eta_{sc} = 0.38, \quad \eta_{sb} = 0.25, \quad \eta_{cd} = 0.24, \quad \eta_{ub} = 0.16.
\end{align}
The resulting quark dressing functions $A(p)$ and $M(p)$ are the same as those computed in Fig.~(5) of Ref.~\cite{Gao:2024gdj}. Here, it is important to remark that when employing these fixed values for $\eta_{f f'}$ the light mesons, together with the heavy light $D$ and $D_s$, can be computed numerically onshell. On the other hand, for the $B, B_s$ and $B_c$ we employed a parametrization of the quark propagator in terms of complex conjugate poles, as described in \cite{Gao:2024gdj}.  

(${\it ii}$) \textbf{BS amplitudes.} To compute the BS amplitude from \1eq{eq:BSEI}, we employ standard methods where the integral equation is reformulated as an eigenvalue problem. The physical solutions correspond to mass-shell points given by $P^2 = -M^2$. In our numerical implementation, the BS amplitude is expanded using a basis of 8 Chebyshev polynomials. Figure~\ref{fig:BSA_amps} displays the resulting subamplitudes $\chi_i^{f f'}$ for the pion and the $D_s$ meson, projected onto the leading Chebyshev moment.

\begin{figure*}[t]
\centerline{%
\includegraphics[width=1.15\textwidth]{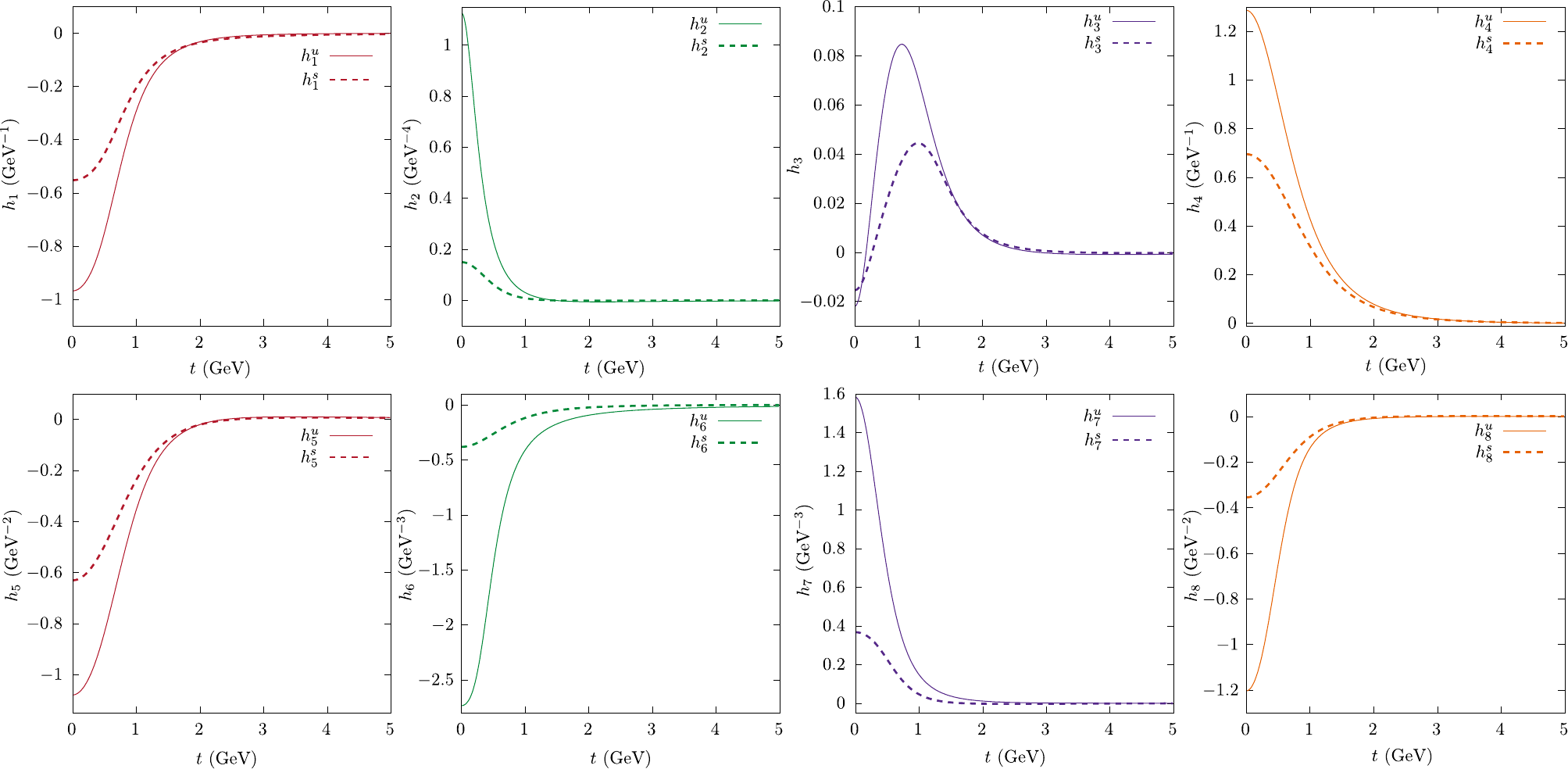}}
\caption{Transverse form factors of the quark-photon vertex  using the interaction from \1eq{eq:Cfun} for up quark (solid lines) and strange quark (dashed lines).}
\label{fig:qphv_dress_trans}     
\end{figure*}

(${\it iii}$) \textbf{Form factors for the quark-photon vertex.} For the calculation of spacelike hadron form factors, the quark-photon vertex is required in the region $q^2 > 0$. The solution of \1eq{eq:vecvert}, unlike the homogeneous case, does not correspond to an eigenvalue problem. Instead, the equation is a linear integral equation driven by the bare vertex, which after discretization reduces to a system of linear equations where the interaction kernel defines the matrix structure. Such systems can be solved using ``LU decomposition", factorizing the kernel matrix into lower and upper triangular parts for efficient inversion \cite{press_etal:1992}. In \Cref{fig:qphv_dress_long} and \Cref{fig:qphv_dress_trans} we present the full structure of the vertex for $u$ and $s$ quarks. As a consistency check, we have verified that our numerical results from the WTI solution from \1eq{eq:Ball_Chiu_FF} are identical to those obtained from the vertex SDE. On the other hand, the transverse components are consistent with previous studies \cite{Miramontes:2019mco, Sanchis-Alepuz:2017jjd}. 
\begin{figure*}[t]
\centerline{%
\includegraphics[width=1.0\textwidth]{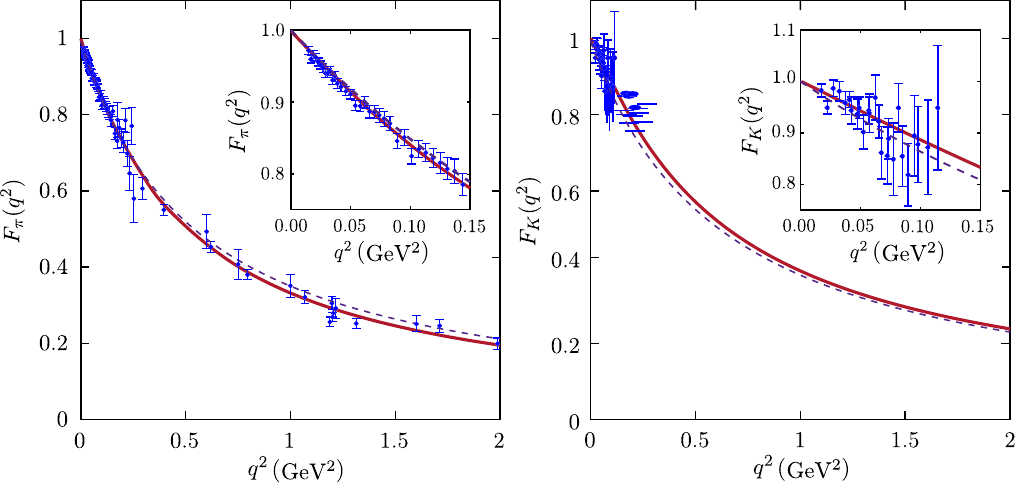}}
\caption{Computed pion (left) and kaon (right) electromagnetic form factor for a space-like photon (solid lines), compared with the available experimental 
data~\cite{Brown:1973wr,Bebek:1974iz,JeffersonLabFpi-2:2006ysh,NA7:1986vav,Dally:1980dj,JeffersonLabFpi:2000nlc} and with the results from the weighted-RL \cite{Xu:2024fun} (dashed-line). The insets offer a closer look at the infrared region of momenta, where most of the experimental points are accumulated.}
\label{fig:pionFF}     
\end{figure*}

\begin{figure*}[t!]
\centerline{%
\includegraphics[width=0.9\textwidth]{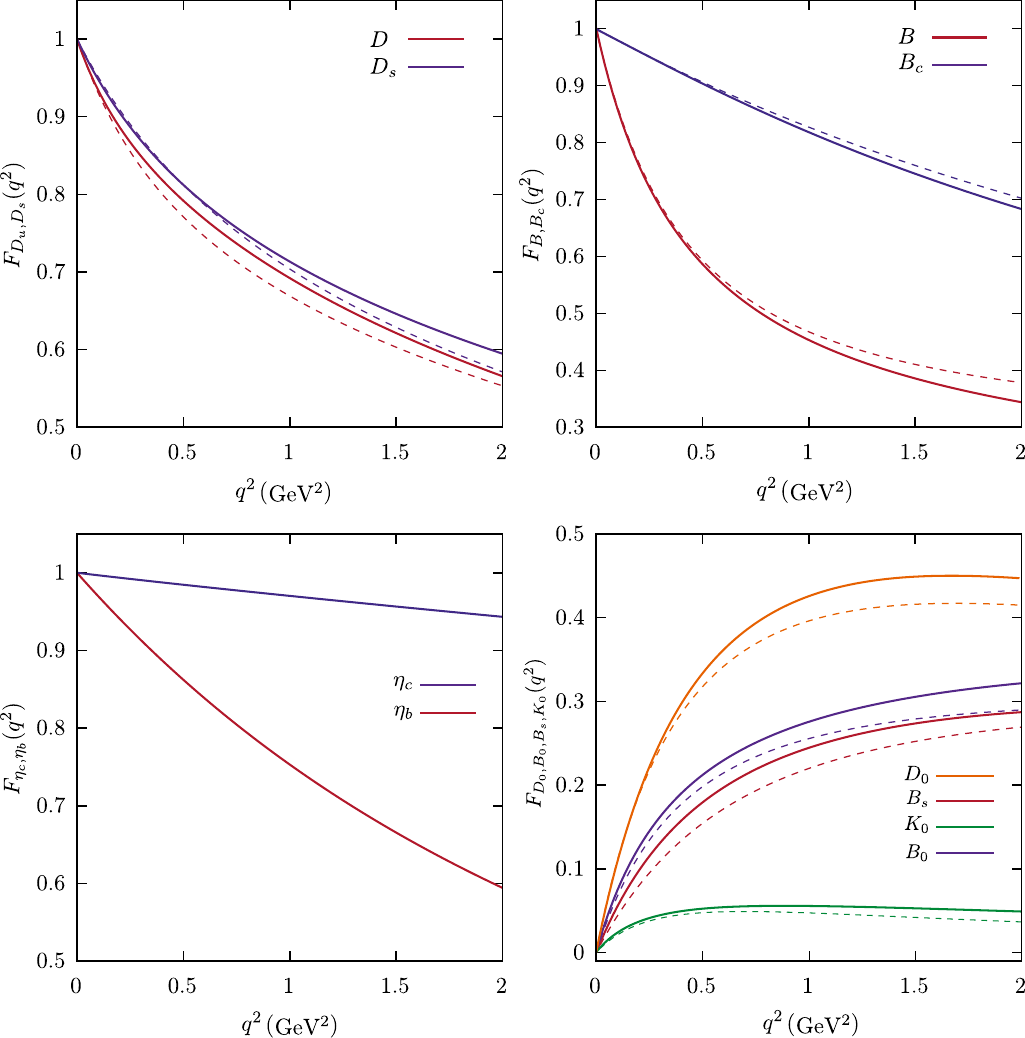}}
\caption{Computed electromagnetic form factors (solid lines) for mesons interacting with a space-like photon compared with the results from \cite{Xu:2024fun} (dashed-line). Results are shown for: $D$ and $D_s$ (upper left), $B$ and $B_c$ (upper-right), $\eta_c$ and $\eta_b$ (bottom-left), and for $K_0, B_0$ and $B_s$ (bottom-right).}
\label{fig:DFF}     
\end{figure*}

(${\it iv}$) \textbf{Electromagnetic form factors.} Moving to the EFF in \fig{fig:pionFF} and \fig{fig:DFF}, we note that our results are correctly normalized, satisfying $F(0) = 1$ for electrically charged particles, and $F(0) = 0$ for neutral mesons, as required by the vector WTI.

The left panel of \fig{fig:pionFF} shows that our result for the pion electromagnetic form factor is in excellent agreement with experimental measurements. Similarly, the right panel of \fig{fig:pionFF} shows that our computation for the kaon electromagnetic form factor is consistent with the available experimental data. Furthermore, our results for the $D$ and $D_s$ mesons exhibit a similar qualitative behavior to that observed between the pion and kaon, see \fig{fig:DFF}, where the inclusion of a strange quark in the meson’s quark content leads to a slight difference between the form factor of $D (c\bar{d})$ and $D_s (c\bar{s})$, reflecting the influence of the quark mass. A similar trend is evident in the comparison between the  $B(u\bar{b})$ and $B_c(c\bar{b})$ form factors, where the presence of a lighter quark enhances the mass asymmetry, resulting in a more pronounced variation in the form factor behavior. In the case of the $\eta_c$ and $\eta_b$ mesons, 
 when both quark and antiquark contributions are included, 
the total electromagnetic form factor vanishes identically due to 
charge conjugation.
Nevertheless, structural information can still be extracted by computing the form factor associated with a single quark contribution from \1eq{eq:FF_full}.  Our results indicate that the electromagnetic form factor for the  $\eta_b$ exhibits a noticeably slower decrease with increasing momentum transfer compared to the $\eta_c$.

(${\it v}$) \textbf{Charge radii.} The charge radius can be determined from the derivative of the electromagnetic form factor evaluated at $q^2 = 0$:
\begin{align}
    \langle r^2 \rangle = -6 \frac{d F(q^2)}{d q^2} \Big \vert_{q^2=0}
    \,.
\end{align}
The computed charge radii for different pseudoscalar mesons are presented in Table \ref{tab:radii}, where they are compared with experimental measurements, as well as predictions from the Contact Interaction (CI) model \cite{Hernandez-Pinto:2023yin}, the Algebraic model (AM) \cite{Almeida-Zamora:2023bqb}, Lattice QCD (LQCD) \cite{Wang:2020nbf, Can:2012tx}, Data driven \cite{Stamen:2022uqh, Leplumey:2025kvv}, Light front \cite{Xu:2025ntz}, and the  
weighted-RL approach of \cite{Xu:2024fun}. 

Our results for the $\pi^+$, $K^+$, and $K^0$ mesons are in very good agreement with experimental values, showing small deviations of 0.45\%, 1.42\%, and 2.5\%, respectively.

Comparing our results with those from other approaches, we observe some differences. The CI model underestimates the charge radii of the pion and kaon, predicting values of 0.450 fm and 0.420 fm, respectively, which are significantly smaller than both experimental measurements and our own results. This underestimation is also evident in the predictions for heavy-light mesons, where, for instance, the $D_s$ meson radius is given as 0.260 fm, notably lower than our 0.368 fm. Nevertheless, the overall qualitative behavior is consistent with our findings.

The Algebraic model yields a charge radius of 0.680 fm for the $D$ meson, which is substantially larger than both our prediction (0.428 fm) and lattice QCD results (0.450 fm). Additionally, this model exhibits a significant difference between the $D$ and $D_s$ mesons, with a predicted charge radius of 0.372 fm for the latter, whereas our results suggest that these two mesons should have comparable charge radii.

For heavy-light mesons, our results exhibit a similar overall behavior compared to the weighted-RL investigated in \cite{Xu:2024fun}. In particular, our predictions for the $B$ and $B_s$ mesons, 0.631 fm and 0.330$i$ fm, respectively, are in close agreement with the corresponding weighted-RL values of 0.619 fm and 0.337$i$ fm. The most noticeable discrepancies arise in the light meson sector, where some variations in the charge radii are observed.

Finally, it is interesting to note that lattice QCD is the only approach that obtains a larger  
central value for the charge radius of the $D_s$ meson than for the $D$ meson. Nevertheless, the error for both particles is rather considerable. 

\begin{table*}[t!]
    \centering
    \footnotesize
    \sisetup{table-format=1.6(2)}
    \renewcommand{\arraystretch}{1.0}
    \rowcolors{6}{gray!25}{white}
    \setlength{\tabcolsep}{3.0pt} 
    \arrayrulecolor{black!40} 
    \begin{tabular}{|l|*{12}{S[table-format=1.3]|}} 
        \hline
        \rowcolor{white}
        {Radius (fm)} & {$\pi$} & {$K$} & {$K^0$} & {$D$} & {$D_0$} &{$D_s$} & {$B$} & {$B_0$} & {$B_s$} & {$B_c$} & {$\eta_c$}  & {$\eta_b$} \\
        \hline
        {This work} & {0.656} & {0.568} & {0.270$i$} & {0.428} & {0.542$i$} &{0.368} & {0.631} & {0.442$i$} &{0.330$i$} & {0.213} & {0.267} & {0.082} \\
        \hline
        {Experiment \cite{ParticleDataGroup:2022pth}}  & {0.659} & {0.560} & {0.277$i$} & {--} & {--} & {--} &{--} & {--} &{--} & {--} & {--} & {--} \\
 \hhline{|=|=|=|=|=|=|=|=|=|=|=|=|=|}
        \rowcolor{white}
         {LQCD \cite{Wang:2020nbf,Can:2012tx}}  & {0.656} & {--} & {--} & {0.450(24)} & {--} & {0.465(57)} &{--} &{--} & {--} & {--} & {--} & {--} \\
        \hline
        \rowcolor{white}
         {Data driven \cite{Stamen:2022uqh,Leplumey:2025kvv}}   & {0.655} & {0.599} & {0.245$i$} & {--} & {--} & {--} &{--} &{--} & {--} & {--} & {--} & {--} \\
        \hline
        \rowcolor{white}
         {Weighted-RL \cite{Xu:2024fun}}  & {0.646} & {0.608} & {0.253$i$} & {0.435} & {0.556$i$} &{0.352} &{0.619} &{0.435$i$} & {0.337$i$} & {0.219} & {--} & {--} \\
         \hline
         \rowcolor{white}
         {Light front \cite{Xu:2025ntz}}  & {0.668} & {0.610} & {0.302$i$} & {0.411} & {0.534$i$} & {0.301} &{0.564} &{0.396$i$} & {0.281$i$} & {0.189} & {--} & {--} \\
         \hline
         \rowcolor{white}
          {AM \cite{Almeida-Zamora:2023bqb}}  & {--} & {--} & {--} & {0.680} & {} & {0.372} &{0.926} &{--} & {0.345$i$} & {0.217} & {--} & {--} \\
        \hline
        \rowcolor{white}
         {CI \cite{Hernandez-Pinto:2023yin}}  & {0.45} & {0.42} & {--} & {--} & {--} & {0.26} &{0.34} &{0.36$i$} & {0.24$i$} & {0.17} & {0.20} & {0.07} \\
        \hline
    \end{tabular}
    \caption{Charge radius predictions for the different pseudoscalar particles compared with the contact interaction model, algebraic model, lattice QCD, Data driven methods and weighted-RL data available. All results are given in fm. Note that the imaginary factor ``$i$" has been used when the 
    squares of the reported radii are  negative.}
    \label{tab:radii}
\end{table*}
%

\section{Conclusions}
\label{sec:conc}

In this work, we have applied the Bethe–Salpeter framework to the computation of EFF of heavy–light pseudoscalar mesons by implementing a flavour–dependent interaction kernel. The computed EFF for light mesons ($\pi$ and $K$) shows excellent agreement with experimental data, thereby validating the consistency of the approach. In the heavy–light sector, where experimental and lattice QCD results for form factors are not yet available, we provide predictions. In particular, our estimates of the charge radii are in good agreement with existing theoretical studies, offering valuable benchmarks for forthcoming lattice and experimental efforts.

Furthermore, the present framework can be naturally extended to the study of vector mesons, such as the $\rho$ and $K^*$, for which EFF provide complementary insight into internal meson structure beyond spin-zero systems. Beyond mesons, the formalism can also be applicable to baryons, which will be investigated in the future.

\section*{Acknowledgements}

The authors thank Yin-Zhen Xu for useful communications.
The work of A.S.M.~and 
J.P.~is funded by the Spanish MICINN grants PID2020-113334GB-I00 and
PID2023-151418NB-I00,  
the Generalitat Valenciana grant CIPROM/2022/66,
and CEX2023-001292-S by MCIU/AEI.
J.P.~is supported 
in part by the EMMI visiting grant of 
the ExtreMe Matter Institute EMMI
at the GSI,
Darmstadt, Germany.
J.M.P.~is  funded by  the Deutsche Forschungsgemeinschaft (DFG, German Research Foundation) under Germany’s Excellence Strategy EXC 2181/1 - 390900948 (the Heidelberg STRUCTURES Excellence Cluster) and the Collaborative Research Centre SFB 1225 - 273811115 (ISOQUANT).

\clearpage
\bibliography{main}

\end{document}